\documentclass{myaa}

\usepackage[varg]{txfonts}

\usepackage{natbib}
\usepackage{graphicx}
\usepackage{epstopdf}
\usepackage{amssymb}
\usepackage{float}
\begin{document}

\title{Model of a protoplanetary disk forming in-situ the major Uranian satellites before the planet is formed}
%%%  $\boldsymbol{G}$

\author{Dimitris M. Christodoulou\inst{1,2}  
\and 
Demosthenes Kazanas\inst{3}
}

%%%$^{2}$\footnotemark[1]
\institute{
Lowell Center for Space Science and Technology, University of Massachusetts Lowell, Lowell, MA, 01854, USA.\\
\and
Dept. of Mathematical Sciences, Univ. of Massachusetts Lowell, 
Lowell, MA, 01854, USA. \\ E-mail: dimitris\_christodoulou@uml.edu\\
\and
NASA/GSFC, Laboratory for High-Energy Astrophysics, Code 663, Greenbelt, MD 20771, USA. \\ E-mail: demos.kazanas@nasa.gov \\
}

%\date{Received~~2019 month day; accepted~~2019~~month day}

\def\gsim{\mathrel{\raise.5ex\hbox{$>$}\mkern-14mu
                \lower0.6ex\hbox{$\sim$}}}

\def\lsim{\mathrel{\raise.3ex\hbox{$<$}\mkern-14mu
               \lower0.6ex\hbox{$\sim$}}}

\abstract{
We fit an isothermal oscillatory density model of Uranus' protoplanetary disk to the present-day major satellites and we determine the radial scale length of the disk, the equation of state and the central density of the primordial gas, and the rotational state of the Uranian nebula. This disk does not at all look like the Jovian disk that we modeled previously. Its rotation parameter that measures centrifugal support against self-gravity is a lot smaller ($\beta_0=0.00507$), as is the radial scale length (only 27.6 km) and the size of the disk (only 0.60 Gm). On the other hand, the central density of the compact Uranian core is higher by a factor of 180 and its core's angular velocity is about 2.3 times that of Jupiter's core (a rotation period of 3.0 d as opposed to 6.8 d). Yet, the rotation of the disk is sufficiently slow to guarantee its long-term stability against self-gravity induced instabilities for millions of years.}

\authorrunning{ }
\titlerunning{Formation of the major Uranian satellites}

\maketitle

%% From the front matter, we move on to the body of the paper.
%% In the first two sections, notice the use of the natbib \citep
%% and \citet commands to identify citations.  The citations are
%% tied to the reference list via symbolic KEYs. The KEY corresponds
%% to the KEY in the \bibitem in the reference list below. We have
%% chosen the first three characters of the first author's name plus
%% the last two numeral of the year of publication as our KEY for
%% each reference.

\section{Introduction}\label{intro}

In previous work \citep{chr19a,chr19b}, we presented isothermal models of the solar and the Jovian primordial nebulae capable of forming protoplanets and protosatellites, respectively, long before the central object is actually formed by accretion processes. This entirely new ``bottom-up'' formation scenario is currently observed in real time by the latest high-resolution ($\sim$1-5~AU) observations of many protostellar disks by the ALMA telescope \citep{alm15,and16,rua17,lee17,lee18,mac18,ave18,cla18,kep18,guz18,ise18,zha18,dul18,fav18,har18,hua18,per18,kud18,lon18,pin18,vdm19}.   In this work, we apply the same model to Uranus' primordial disk that formed its six major satellites. Our goal is to compare our best-fit model of Uranus' primordial nebula to Jupiter's nebula and to find similarities and differences between the two disks that hosted gravitational potential minima in which the orbiting moons could form in relative safety over millions of years of evolution.

As was expected, the two model nebulae are very different in their radial scale lengths (27.6 km versus 368 km, for Uranus and Jupiter, respectively) and their sizes (0.60 Gm versus 12 Gm, respectively) and central densities (55.6 g~cm$^{-3}$ versus 0.31 g~cm$^{-3}$, respectively). In addition to structural differences, the disks are significantly different in their remaining physical quantities: Uranus' core is smaller by about a factor of 2 ($R_1\approx 0.1$ Gm), the radial density profile is shallower ($k\approx -1$), and there is no need for an outer flat-density region; also, Uranus' disk enjoys a lot lower rotational support against self-gravity than Jupiter's disk ($\beta_0\approx 5\times 10^{-3}$ versus $\beta_0\approx 3\times 10^{-2}$, respectively).

The extremely high gas densities and the mild differential rotation speeds in Uranus' compact disk signify that its major equatorial moons were formed long before the planet was actually fully formed; but not before the protoplanet was knocked over to its current axial tilt of 98 degrees. This is because all of the inner moons orbit at nearly zero inclination to the planet's equator and this means that the accretion disk was formed around the protoplanetary core after it had been tilted severely by a very early giant impact.

The analytic (intrinsic) and numerical (oscillatory) solutions of the isothermal Lane-Emden equation and the resulting model of the gaseous nebula have been described in detail in \cite{chr19b} for Jupiter's disk, and there is no need of repeating the descriptions here. 
In what follows, we apply in \S~\ref{models2} our model nebula to the major moons of Uranus and we compare the best-fit results to Jupiter's extended {\it Model 2}. In \S~\ref{disc}, we summarize and discuss our results.

\begin{figure}
\begin{center}
    \leavevmode
      \includegraphics[trim=0.2 0.2cm 0.2 0.2cm, clip, angle=0, width=10 cm]{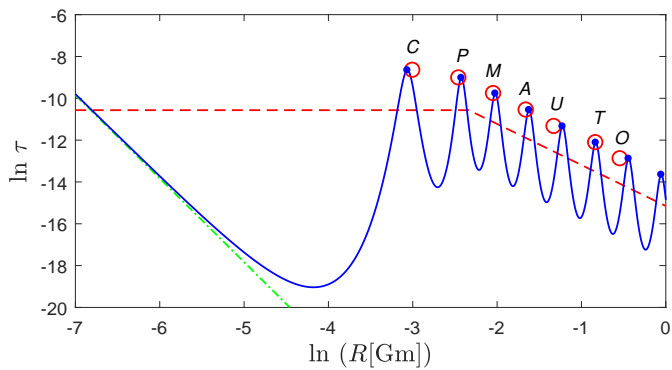}
      \caption{Equilibrium density profile for the midplane of Uranus' primordial protoplanetary disk that formed its largest moons. The innermost small moon Cordelia was also retained because it improves the fit. (Key: C:Cordelia, P:Puck, M:Miranda, A:Ariel, U:Umbriel, T:Titania, O:Oberon.) The best-fit parameters are $k=-0.96$, and $\beta_0=0.00507$ (or, equivalently, $R_1=0.0967$~Gm). The radial scale length of the disk is only $R_0=27.6$~km. The Cauchy solution (solid line) has been fitted to the present-day moons of Uranus so that its density maxima (dots) correspond to the observed semimajor axes of the orbits of the moons (open circles). The density maximum corresponding to the location of Titania was scaled to a distance of $R_T=0.4358$~Gm. The mean relative error of the fit is 7.5\%, affirming that this simple equilibrium model produces a good match to the observed data points. The intrinsic solution (dashed line) and the nonrotating analytical solution (dash-dotted line) are also shown for reference. 
\label{fig1}}
  \end{center}
  %\vspace{-1.35cm}
\end{figure}

\section{Physical Model of Uranus' Protoplanetary Disk}\label{models2}

\subsection{Best-Fit Uranian disk model}\label{model1}

The numerical integrations that produce oscillatory density profiles were performed with the \textsc{Matlab} {\tt ode15s} integrator \citep{sha97,sha99} and the optimization used the Nelder-Mead simplex algorithm as implemented by \cite{lag98}. This method (\textsc{Matlab} routine {\tt fminsearch}) does not use any numerical or analytical gradients in its search procedure which makes it extremely stable numerically, albeit somewhat slow.

In Fig.~\ref{fig1}, we show the best optimized fit to the semimajor axes of the moons of Uranus. The innermost small moon Cordelia was used along with the other satellites (Puck, Miranda, Ariel, Umbriel, Titania, and Oberon) because its inclusion improves the best fit substantially. The need for including Cordelia stems from the structure of the oscillatory solutions of the Lane-Emden equation \citep{lan69,emd07} with rotation: the solutions track closely the nonrotating solution at small radii and they turn around to approach the intrinsic solution (creating oscillations) only after they cross below the intrinsic core density value of $\tau = \beta_0^2$ \citep{jea14}. Even for extremely low values of the parameter $\beta_0$, these solutions create density maxima deep inside the core region, so there is need for at least one moon residing well inside the core and close to the center $R=0$. 

We have effectively used only two free parameters ($k$ and $\beta_0$) to fit the current orbits of the seven satellites, and the best-fit model is of good quality (mean relative error of 7.5\%, all of which is coming from the positions of Umbriel and Oberon). The inner core parameter $R_1$ is strongly correlated to $\beta_0$, and we did not use an outer flat-density region beyond the radius $R_2$ because the disk is too small in radial extent.

We find the following physical parameters from the best-fit model: $k=-0.96$ and $\beta_0=0.00507$ (equivalently, $R_1=0.0967$ Gm, just beyond the orbit of Puck and closer to the orbit of the minor moon Mab). The radial scale of the model was determined by fitting the density peak that corresponds to the orbit of Titania to its distance of 0.4358 Gm, and the scale length of the disk then turns out to be $R_0=27.6$ km. The best-fit model is certainly stable to nonaxisymmetric self-gravitating instabilities because of the extremely low value of $\beta_0$ \citep[the critical value for the onset of dynamical instabilities is $\beta_*\simeq 0.50$;][]{chr95}.

The model disk extends out to 1 Gm ($\ln R=0$ in Fig.~\ref{fig1}), but its validity ends around the distance of the outermost major moon Oberon ($R_{\rm max}\approx 0.60$ Gm). The next outer density peak lies at a distance of 0.95 Gm around which no moon is known. In fact, the disk of Uranus must have been really small ($< 1$ Gm in radial extent) because the next outer irregular moon, Francisco, has a semimajor axis of 4.3 Gm.

\subsection{Physical parameters from the best-fit model}\label{rhomax1}

Using the scale length of the disk $R_0$ and the definition $R_0^2 = c_0^2/(4\pi G\rho_0)$, we write the equation of state for the Uranian circumplanetary gas as
\begin{equation}
\frac{c_0^2}{\rho_0} \ = \ 4\pi G R_0^2 \ = \ 6.39\times 10^{6} 
{\rm ~cm}^5 {\rm ~g}^{-1} {\rm ~s}^{-2}\, ,
\label{crho1}
\end{equation}
where $c_0$ and $\rho_0$ are the local sound speed and the local density in the inner disk, respectively, and $G$ is the gravitational constant.
For an isothermal gas at temperature $T$, ~$c_0^2 = {\cal R} T/\overline{\mu}$, where $\overline{\mu}$ is the mean molecular weight and ${\cal R}$ is the
universal gas constant. Hence, eq.~(\ref{crho1}) can be rewritten as
\begin{equation}
\rho_0 \ = \ 13.0 \,\left(\frac{T}{\overline{\mu}}\right) \
{\rm ~g} {\rm ~cm}^{-3}\, ,
\label{trho1}
\end{equation}
where $T$ and $\overline{\mu}$ are measured in degrees Kelvin and 
${\rm ~g} {\rm ~mol}^{-1}$, respectively. 

For the coldest gas with $T \geq 10$~K 
and $\overline{\mu} = 2.34 {\rm ~g} {\rm ~mol}^{-1}$ (molecular hydrogen and
neutral helium with fractional abundances $X=0.70$ and $Y=0.28$ by
mass, respectively), we find that
\begin{equation}
\rho_0 \ \geq \ 55.6 \ {\rm ~g} {\rm ~cm}^{-3}\, .
\label{therho1}
\end{equation}
This extremely high value implies that the conditions for protosatellite formation were already in place during the early isothermal phase \citep{toh02} of the Uranian nebula.

Using the above characteristic density $\rho_0$ of the inner disk
in the definition of ~$\Omega_J\equiv\sqrt{2\pi G\rho_0}$, we determine the Jeans frequency of the disk:
\begin{equation}
\Omega_J \ = \ 4.8\times 10^{-3} {\rm ~rad} {\rm ~s}^{-1}\, .
\label{thej1}
\end{equation}
Then, using the model's value $\beta_0 = 0.00507$ in the definition 
of ~$\beta_0\equiv \Omega_0 /\Omega_J$, we determine the angular velocity of the uniformly-rotating core ($R_1\leq 0.0967$~Gm), viz.
\begin{equation}
\Omega_0 \ = \ 2.5\times 10^{-5} {\rm ~rad} {\rm ~s}^{-1}\, .
\label{theom1}
\end{equation}
For reference, this value of $\Omega_0$ for the core of the Uranian nebula corresponds to an orbital period of $P_0=3.0$~d. This value is close to the present-day orbital period of Ariel (2.5 d), but it is not near the orbital period of the largest moon Titania (8.7 d). This is a deviation from what we found for the solar system and for Jupiter. Nevertheless, the large outer moons of Uranus are all comparable in mass and size, so our previous finding remains valid: the angular velocity of the core of the primordial nebula is comparable to the present-day angular velocities of the largest regular satellites.

\begin{table*}
\caption{Comparison of the protoplanetary disks of Jupiter and Uranus}
\label{table1}
\begin{tabular}{llll}
\hline
Property & Property & Jupiter's & Uranus' \\
Name     & Symbol (Unit) & Model 2 & Best-Fit Model \\
\hline
Density power-law index & $k$                                          &   $-1.4$  	     & $-0.96$    \\
Rotational parameter & $\beta_0$                                &    0.0295 	       &  0.00507   \\
Inner core radius & $R_1$ (Gm)                              &   0.220  	       &  0.0967    \\
Outer flat-density radius & $R_2$ (Gm)                              &   5.37        	   &  $\cdots$   \\
Scale length & $R_0$ (km)                               &   368        	   &  27.6   \\
Equation of state & $c_0^2/\rho_0$ (${\rm cm}^5 {\rm ~g}^{-1} {\rm ~s}^{-2}$) & $1.14\times 10^9$ & $6.39\times 10^6$    \\
Minimum core density for $T=10$~K, $\overline{\mu} = 2.34$ & $\rho_0$ (g~cm$^{-3}$)         &    0.31   			&  55.6   \\
Isothermal sound speed for $T=10$~K, $\overline{\mu} = 2.34$ & $c_0$ (m~s$^{-1}$) & 188 & 188 \\
Jeans gravitational frequency & $\Omega_J$ (rad~s$^{-1}$)    &    $3.6\times 10^{-4}$ & $4.8\times 10^{-3}$    \\
Core angular velocity & $\Omega_0$ (rad~s$^{-1}$)    &    $1.1\times 10^{-5}$ 	& $2.5\times 10^{-5}$    \\
Core rotation period & $P_0$ (d)                                 &    6.8 	   			&  3.0   \\
Maximum disk size & $R_{\rm max}$ (Gm)                &    12 	   			&   0.60  \\
\hline
\end{tabular}
\end{table*}

\subsection{Comparison between the models of Uranus and Jupiter}\label{comp}

We show a comparison between the physical parameters of the best-fit models of Uranus and Jupiter in Table~\ref{table1}. Obviously, these two protoplanetary disks are very different in most of their physical properties. The disk of Uranus is a lot smaller ($R_{\rm max}$), more compact ($R_0$), and denser ($\rho_0$) by a factor of 180. In addition, the disk of Uranus is just as cold (assuming that $T=10$ K), a lot heavier ($\Omega_J$), and its core is rotating ($\Omega_0$) more than twice as fast (still, this is a slow rotation with a period $P_0$ of only 3.0 d). 

The power-law index of the Uranian nebular model is $k\approx -1$ (surface density $\Sigma\propto R^{-1}$), unlike the Jovian nebula and the solar nebula ($k\approx -1.5$, $\Sigma\propto R^{-1.5}$). This range of values of $k$ has been observed in studies of young circumstellar disks in the pre-ALMA era \citep[][and references within]{and07,hun10,lee18}.

Although the disk of Uranus is small, it is still very heavy and hosts high densities of gas, thus also of ices and planetesimals. As we found for Jupiter's disk, these conditions support a  ``bottom-up'' hierarchical formation in which protosatellites are seeded early inside such nebular disks and long before their protoplanets are fully formed; these compact moon systems complete their formation in $< 0.1$ Myr \citep{har18} and long before the central stars become fully formed \citep{gre10}.

\section{Summary ans Discussion}\label{disc}

We have constructed isothermal differentially-rotating protoplanetary models of the Uranian nebula, the primordial disk in which the regular moons were formed (\S~\ref{models2}). The best-fit model is shown in Fig.~\ref{fig1} and its physical parameters are listed in Table~\ref{table1}. In the optimization, we retained also the smaller moons Cordelia and Puck in order to fit the nearest two density maxima to the center that form inside the uniform core of the disk. This allowed us to find a better model than when major moons are assumed to have formed inside the core. The remaining mean relative error of 7.5\% was due to inaccuracies in the positions of only two moons, Umbriel and Oberon.

We have compared this model to the best-fit model of Jupiter \citep{chr19b} (\S~\ref{comp}). Uranus' disk has a lot less centrifugal support which makes it extremely stable against dynamical nonaxisymmetric instabilities. Furthermore, this disk is much smaller, heavier, and denser than the disk of Jupiter. Despite these values, Uranus' inner core rotates about twice as fast, although its temperature (as measured by the isothermal sound speed; Table~\ref{table1}) is assumed to be the same as that of the Jovian disk.

Both of these models appear to be stable and long-lived, so their regular moons can form early in the evolution of each nebula and long before the protoplanets manage to pull their gaseous envelopes on to their solid cores. The results support a ``bottom-up'' scenario in which regular satellites form first, followed by their planets, and then by the central star.

Our modeling efforts have produced very different models for Uranus' and Jupiter's disk. This supports the optimization procedure that led to such different results. This model appears to be capable of navigating between varied physical conditions and of finding the best-fit model in each different case. The results for the two gas giants so far indicate that at least two minor planets form inside the uniform core of the model, but the major planets form along the intrinsic density gradient characterized by the power-law index ($k-1$). However, there is still a large region in the inner core where no moons form; this region is where rings form around the gas giants. These rings must also have formed at density maxima, and they may have to be taken into account in future modeling of Saturn and Neptune. Fitting the same model to these gas giants encounters more difficulties than the models of Jupiter and Uranus. We are in the process of investigating the protoplanetary disks of Saturn and Neptune as well.

\iffalse
\section*{Acknowledgments}
%We thank anonymous referees for suggestions that clarified considerably the presentation of these ideas. 
NASA support over the years is gratefully acknowledged.
\fi

%\label{lastpage}

\end{document}